\documentclass[twocolumn,superscriptaddress,floatfix,preprintnumbers,prl,nofootinbib,nobibnotes,amsmath,amssymb,aps]{revtex4-1}
\usepackage{hyperref}

\usepackage{graphicx}
\usepackage{dcolumn}
\usepackage{bm}
\usepackage{color}
\usepackage{xcolor}
\usepackage{soul}

\begin{document}

\preprint{FERMILAB-PUB-23-0832-T\quad\,\,\,\,\quad\quad\quad\quad\quad\quad\quad\quad\quad\quad\quad\quad DESY-24-069 \quad\quad\quad\quad\quad\quad\quad\quad\quad\quad\quad\quad\quad\quad\quad\quad APS/123-QED}

\title{
Discovering Electroweak Interacting Dark Matter at Muon Colliders using Soft Tracks
}
\author{Rodolfo Capdevilla}
\affiliation{Particle Theory Department, Fermi National Accelerator Laboratory  , Batavia, IL 60510, USA}
\email{rcapdevi@fnal.gov}

\author{Federico Meloni}
\affiliation{Deutsches Elektronen-Synchrotron DESY, Notkestr.\ 85, 22607 Hamburg, Germany}
\email{federico.meloni@desy.de}

\author{Jose Zurita}
\affiliation{\it Instituto de F\'{\i}sica Corpuscular, CSIC-Universitat de Val\`encia, Valencia, Spain}
\email{jzurita@ific.uv.es}

\date{\today}

\begin{abstract}
Minimal Dark Matter models feature one neutral particle that serves as a thermal relic dark matter candidate, as well as quasi-degenerate charged states with TeV masses. When the charged states are produced at colliders, they can decay into dark matter and a low-momentum (soft) charged particle, which is challenging to reconstruct at hadron colliders. We demonstrate that a 3 TeV Muon Collider is capable of detecting these soft tracks, enabling the discovery of thermal Higgsinos and similar dark matter candidates which constitute highly motivated scenarios for future collider searches.
\end{abstract}

\maketitle

{\bf Introduction -} The discovery of the Higgs boson and the multiple measurements confirming the predictions of the Standard Model (SM) validate the successful ongoing program at the Large Hadron Collider (LHC). A next step at the precision frontier is to measure the Higgs properties to disentangle the microscopic nature of the electroweak symmetry-breaking mechanism. On another front, the current lack of new particles around the TeV scale motivates the need for a new boost at the energy frontier. Muon colliders (MuC) are strong candidates for explorations at the 10 TeV parton center-of-momentum (CoM) energies. The physics program of future MuC promises both high-precision Higgs and SM measurements~\cite{Han:2020pif,Ruiz:2021tdt,Han:2021lnp,Chen:2022msz,Forslund:2022xjq,Chen:2022yiu,Ruhdorfer:2023uea,Forslund:2023reu,Liu:2023yrb,Stylianou:2023xit,Celada:2023oji,Cassidy:2023lwd,Han:2020uid,Han:2021kes,Garosi:2023bvq} as well as direct production of new particles with masses well above LHC reach~\cite{AlAli:2021let,Aime:2022flm,MuonCollider:2022xlm,Black:2022cth,Accettura:2023ked,Capdevilla:2020qel,Huang:2021nkl,Capdevilla:2021rwo,Liu:2021jyc,Han:2021udl,Asadi:2021gah,Sen:2021fha,Bandyopadhyay:2021pld,Liu:2021akf,Casarsa:2021rud,Capdevilla:2021kcf,Bao:2022onq,Inan:2022rcr,Liu:2022byu,Lv:2022pts,Kwok:2023dck,Inan:2023pva,Jueid:2023zxx,Jueid:2023qcf,Dermisek:2023tgq,Belfkir:2023lot,Dermisek:2023rvv,Asadi:2023csb,Liu:2023jta}.

Weakly Interacting Massive Particles (WIMPs) are well-motivated dark matter (DM) candidates that naturally appear in many extensions of the SM. Seeking these particles is one of the pillars for physics beyond the Standard Model (BSM) at colliders. In typical WIMP models, the DM candidate is usually accompanied by other states that can leave distinctive imprints at colliders. Such is the case of Minimal Dark Matter (MDM) models, where a single electroweak multiplet is added to the SM~\cite{Cirelli:2005uq}. Depending on their quantum numbers, the {\it thermal mass} (mass value to explain the observed relic abundance, $\Omega_{\rm DM} h^2 = 0.1198$~\cite{Planck:2015fie}) of the MDM multiplets falls in the range of 1-200 TeV~\cite{Bottaro:2021snn,Bottaro:2022one}.

The pure singlet case is phenomenologically ruled out~\cite{Arkani-Hamed:2006wnf}; hence, the new multiplet should be at least a weak doublet. This, in turn, means that the dark sector must include at least one electrically charged particle $\chi^{\pm}$ quasi-degenerate with the dark matter candidate $\chi^0$, with mass splittings $\Delta \lesssim$ GeV~\cite{Thomas:1998wy}. We note that the well-studied Higgsino and Wino cases from Minimal Supersymmetric Standard Model correspond to the Majorana fermion doublet and triplet, respectively. Furthermore, large multiplet MDM models accounting for only a fraction ($1-10\%$) of the relic abundance also have TeV scale masses and are a well-motivated target. Other quasi-degenerate WIMP scenarios include slepton-like particles, extended scalar sectors, and heavy neutral leptons~\cite{Barr:2015eva,Belanger:2018sti,Jana:2020qzn,Chiang:2020rcv,Calibbi:2021fld}.

\begin{figure}
    \centering
\includegraphics[width=0.9\linewidth]{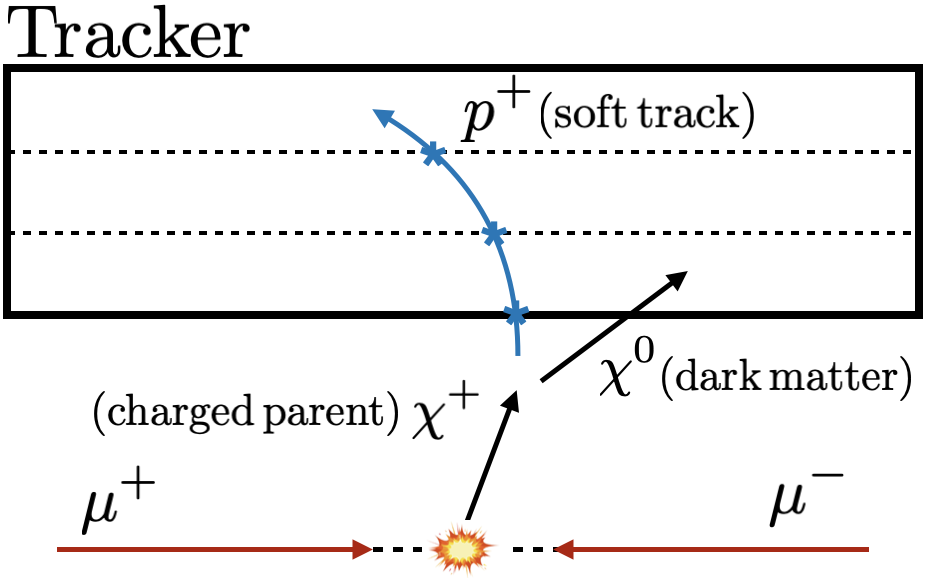}
    \caption{Soft tracks are originated when a heavy particle decays into an almost mass-degenerate dark matter particle $\chi^{0}$ and a charged particle $p^{+}$, which can be reconstructed as a $p_T\lesssim 1$~GeV track.}
    \label{fig:sketch}
\end{figure}

The decay process $\chi^{\pm} \to \chi^0 p^{\pm}$, where $p^{\pm}$ denotes a low transverse momentum or {\it soft} charged particle (Fig.~\ref{fig:sketch}) serves as the primary signature of MDM models. Phenomenological investigations aimed at discovering these models involve the use of monojet \cite{Schwaller:2013baa,Baer:2014cua,Agin:2023yoq} or soft-lepton searches suitable for $\Delta \gtrsim \text{GeV}$~\cite{Han:2014kaa,Low:2014cba}, and disappearing tracks (DT) targeting $\Delta \lesssim \text{GeV}$~\cite{Mahbubani:2017gjh,Fukuda:2017jmk,Bharucha:2018pfu,Calibbi:2018fqf,Filimonova:2018qdc,Belyaev:2020wok} and have sparked an extensive search program at the LHC~\cite{Canepa:2020ntc,ATLAS:2013ikw,CMS:2014gxa,ATLAS:2017oal,CMS:2020atg,ATLAS:2022rme,CMS:2015epp,ATLAS:2017vat,CMS:2018kag,ATLAS:2019lng,ATLAS:2021moa,CMS:2021cox,CMS:2021edw,ATLAS:2021kxv,CMS:2021far}. Prospects for direct searches for MDM at future colliders have been investigated at FCC-hh~\cite{Han:2018wus,Saito:2019rtg}, CLIC~\cite{CLIC:2018fvx}, and the MuC~\cite{Han:2020uak,Capdevilla:2021fmj,Bottaro:2021srh,Bottaro:2021snn,Bottaro:2022one,Black:2022qlg,Bandyopadhyay:2024plc}. Although most of these searches primarily target DT of the charged parent $\chi^\pm$, proposals have also been made to search for the soft charged particle $p^\pm$ at electron-proton colliders~\cite{Curtin:2017bxr} (as well as the LHC~\cite{Fukuda:2019kbp,ATLAS:2019pjd}). Furthermore, indirect searches at MuC \cite{Franceschini:2022sxc,Fukuda:2023yui} provide
a complementary probe.

Proposed searches for MDM models suggest that both a 100~TeV proton machine, such as the FCC-hh, and a 10~TeV muon collider (MuC10), will have the capability to discover the ``Wino-like'' state with a thermal mass of 2.7~TeV. The prospect of discovering the ``Higgsino-like'' state with a thermal mass of 1.1~TeV remains uncertain (some studies suggest that MuC10 may achieve this feat~\cite{Han:2020uak}). Larger multiplets with thermal masses beyond the reach of any foreseeable collider, such as the five-plet with masses around 10-15 TeV~\cite{Bottaro:2021snn,Bottaro:2022one}, would necessitate colliders with CoM energies exceeding 30 TeV~\cite{Han:2020uak}, or indirect probes~\cite{Franceschini:2022sxc,Fukuda:2023yui}.

\begin{figure}[t]
    \centering
\includegraphics[width=0.48\linewidth]{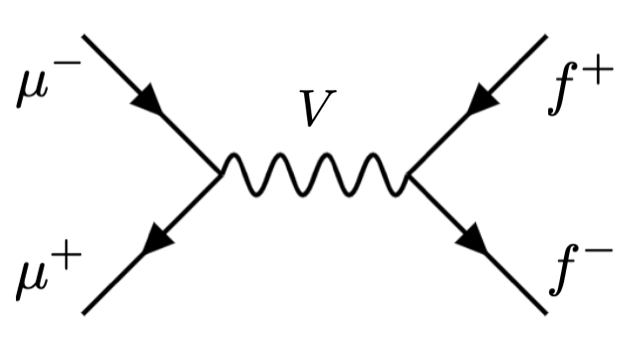}
~~\includegraphics[width=0.48\linewidth]{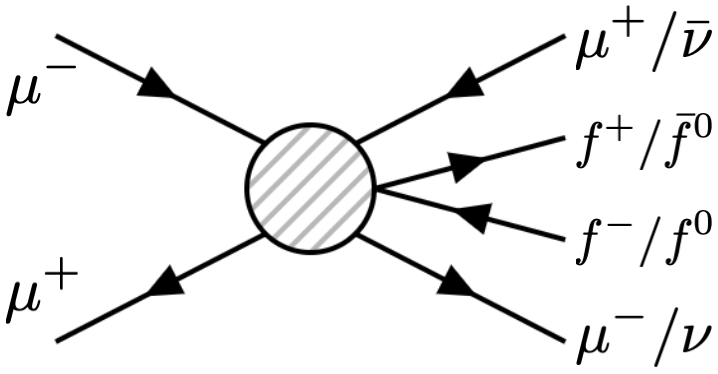}
    \caption{Signal ($f^\pm=\chi^\pm$, $f^0=\chi^0$) and main backgrounds ($f^\pm=\ell^\pm,\tau^\pm,j$, $f^0=\nu_\ell,\nu_\tau,j$). The left diagram is the Drell-Yan-like process, whereas the right blob represents either Vector Boson Fusion or Bhabha scattering with extra emission of a vector boson that then splits into a pair of fermions.}
    \label{fig:FD}
\end{figure}

In this letter, we demonstrate how a dedicated search for the soft charged particle $p^\pm$ can serve as the discovery channel for MDM models, even at the early stage of the Muon Collider program with a CoM energy of 3~TeV (MuC3). Our Soft Track (ST) analysis, indicates that MuC could discover the Higgsino-like thermal relic. Our analysis utilize MuC detector simulations and a rigorous (conservative) treatment of the beam-induced background (BIB) from the decay of the muons as they travel through the machine lattice. As a bonus, our ST analysis can provide unique insights into the dark sector by probing the mass splittings between the DM and its companions, potentially distinguishing MDM from other scenarios. This result highlights the importance of MuC3 not only as a stage towards MuC10 but also as a powerful discovery machine.

{\bf Setup -} We consider a simplified model composed of an electrically charged particle, $\chi^{\pm}$, and a neutral one, $\chi^0$, with a small mass gap, $\Delta = m_{\chi^{\pm}} - m_{\chi^0} \ll m_{\chi^{\pm},\chi^{0}} \equiv m_{\chi}$. As a proof of concept, in this letter we focus on a search for a Higgsino-like state at the 3 TeV MuC (hence we will refer to $\chi^{\pm}$ and $\chi^{0}$ as charginos and neutralinos respectively).

Our focus on this benchmark model is twofold: on one hand, this is one of the most challenging cases for MDM searches at future colliders. On the other hand, its thermal mass (1.1 TeV) is low enough for it to be reachable at a MuC3. To go beyond our benchmark, in our last section we will indicate how to recast our results to other models.

The $\chi$ particles are odd under a $Z_2$ symmetry that renders $\chi^0$ stable (hence a good dark matter candidate) and forces these particles to be pair-produced at colliders. Fig.~\ref{fig:FD} illustrates the main production modes for both signal and background processes in our analysis. Signal production, for which $f^\pm=\chi^\pm$ and $f^0=\chi^0$, is dominated by Drell-Yan-like process (DY).
Background production, for which $f^\pm=\ell^\pm,\tau^\pm,j$, where $\ell^\pm$ stands for an electron or muon, and $f^0=\nu_\ell,\nu_\tau,j'$, is dominated by two processes: Vector Boson Fusion (VBF), where the initial state muons radiate vector bosons that interact, producing the desired final state, and $t$-channel Bhabha-like Scattering (BS) with the extra emission of a vector boson splitting into a pair of fermions.

\begin{table}[t!]
\renewcommand{\arraystretch}{1.25}
\begin{tabular}{|lll|}
\hline 
\multicolumn{3}{|c|}{$\mu^+ \mu^- \to \gamma + {\rm X} \,\,(+\, Z\to\nu\nu)$ }                                                                  \\ \hline
\multicolumn{1}{|l|}{X} & \multicolumn{1}{c|}{$\,\,\,\,\sigma(\gamma {\rm X})\,$ {[}fb{]}$\,\,\,\,$} & \multicolumn{1}{c|}{{$\,\,\,\,\sigma(\gamma {\rm X} Z)\,$ {[}fb{]}$\,\,\,\,$}} \\
\hline
\multicolumn{1}{|l|}{$\ell^+\ell^-\nu_\ell\bar{\nu}_\ell$}          & \multicolumn{1}{c|}{242.0} & \multicolumn{1}{c|}{2.828} \\
\multicolumn{1}{|l|}{$\ell^+\ell^-\mu^+\mu^-$}                & \multicolumn{1}{c|}{60.45} & \multicolumn{1}{c|}{0.012} \\
\multicolumn{1}{|l|}{$e^+\nu_e\mu^-\bar{\nu}_\mu$ + CP \,\,}  & \multicolumn{1}{c|}{226.6} & \multicolumn{1}{c|}{2.710} \\
\hline
\multicolumn{1}{|l|}{$\tau^+\tau^-\nu_\ell\bar{\nu}_\ell$}          & \multicolumn{1}{c|}{6.493} & \multicolumn{1}{c|}{0.058} \\
\multicolumn{1}{|l|}{$\tau^+\tau^-\mu^+\mu^-$}                & \multicolumn{1}{c|}{30.86} & \multicolumn{1}{c|}{0.006} \\
\multicolumn{1}{|l|}{$\tau^+\nu_\tau\mu^-\bar{\nu}_\mu$ + CP} & \multicolumn{1}{c|}{226.2} & \multicolumn{1}{c|}{2.722} \\
\hline
\multicolumn{1}{|l|}{$jj\nu_\ell\bar{\nu}_\ell$}                    & \multicolumn{1}{c|}{104.5} & \multicolumn{1}{c|}{0.904} \\
\multicolumn{1}{|l|}{$jj\mu^+\mu^-$}                          & \multicolumn{1}{c|}{30.63} & \multicolumn{1}{c|}{0.019} \\
\multicolumn{1}{|l|}{$jj\mu^-\bar{\nu}_\mu$ + CP}             & \multicolumn{1}{c|}{1215.} & \multicolumn{1}{c|}{11.57} \\
\hline
\end{tabular}
\caption{Background processes for the 3 TeV MuC. The third column corresponds to extra emission of a $Z$ boson that decays invisibly. The symbol ``+~CP" indicates that the charge-conjugated process is included. These results were obtained using the cuts $p_T^\gamma\ge20$ GeV, $|\eta_\gamma|<2.44$, $p_T^\ell\ge0.1$ GeV, $p_T^j\ge0.1$ GeV, $p_T^\ell\ge0.1$ GeV and no cuts on $\eta_\ell$ and $\eta_j$.}
\label{tab:bgds}
\end{table}

{\bf Signal -} The expected total thermal Higgsino production cross section at $\sqrt{s}=3$~TeV is 12.53 fb. 

The decay $\chi^{\pm} \to \chi^0 p^{\pm}$ gives rise to a ST and large missing energy.
The distinctive signature we pursue involves a soft charged particle, produced near the interaction point (IP), which traverses most layers of the tracker. Objects that can meet these requirements include $\ell,\tau$ or hadrons. In the case of thermal Higgsinos, the charged state has a 90\% branching ratio (BR) to pions, with only a 10\% BR to light leptons. The characteristic transverse momentum of our ST is approximately $p_T \sim \gamma \Delta$ in the laboratory frame, where $\Delta \sim 0.354$ GeV and $\gamma$ is the boost of the parent chargino from which the ST originates. At MuC3, a pair of 1.1 TeV charginos is predominantly produced with a low boost. This implies a $p_T$ spectrum that peaks around few hundreds of MeV with no events beyond 1 GeV. This challenging signature is the focus of our study.

Our signal region of interest includes 2 STs with a polar angle $\theta \in (\pi/3, 2\pi/3)$ and at least one hard photon with $p_T^\gamma > 40$~GeV.
The two STs must have opposite electric charge and a $p_T \in (0.2, 0.75)$~GeV. The photon cut is pivotal for effective signal-to-background discrimination. In order to balance the large $p_T^\gamma$, the charged particles in the background processes acquire a substantial $p_T$. However, the signal processes are largely unaffected: the impact on the ST spectrum is minimal because the considerable transverse energy in the event is balanced by the neutralinos. Finally, we require the ST to have an angular distance of $\Delta R > 1.5$ to help eliminate background from soft photons splitting into light fermions.

{\bf Background -}
The cross sections of the main background processes are listed in Table~\ref{tab:bgds}. These correspond to $\mu^+\mu^-\to\gamma$ X, where X is one of the final states in the left column. 
The ``+ CP" mark indicates that the quoted cross section includes the charge-conjugated process. Very soft photons splitting into a pair of light leptons is a particularly large source of background. We can see this by comparing how the process $\ell^+\ell^-\nu_\ell\bar\nu_\ell$ has a much larger cross section than the analogous processes with taus or quarks. (Our analysis includes processes with full collinear initial state photons, not shown in Table \ref{tab:bgds}, simulated with the improved Weizsacker-Williams approximation implemented in MadGraph \cite{Frixione:1993yw}. These backgrounds are easily removed by our $p_T^\gamma$ cut.)

For all final states in the left column, we also considered an additional $Z \to \nu \nu$ emission, which can alter the transverse kinematics and thus pollute our SR. Emitting this extra particle significantly reduces the cross section. However, some processes have a cross section comparable to the signal even after the emission of the extra $Z$ boson. The reported values include the invisible branching ratio penalty for the $Z$ boson. We note that adding background processes with an extra invisibly-decaying $Z$ boson adds the complication of increasing the particle multiplicity in the final state. This can slow down the calculation beyond a reasonable point. In order to speed up the calculation of said processes, we use the ``mixed beam'' (MB) approximation, where one of the incoming muon beams is replaced by an electron~\cite{Ruiz:2021tdt}. This removes in a gauge invariant manner diagrams with neutral current that do not correspond to VBF topologies and contribute less than $1\%$ of the total cross section.

\begin{figure}[t]
    \centering
\includegraphics[width=0.94\linewidth]{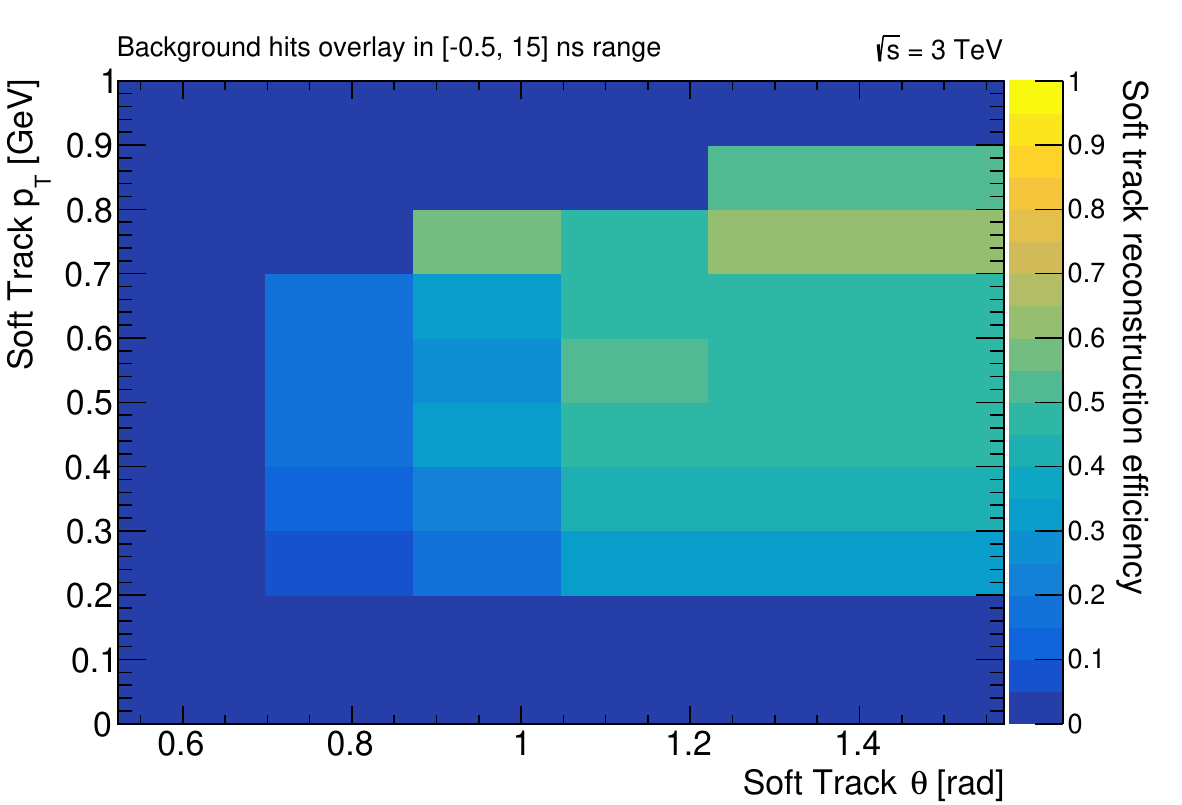}
    \caption{Track reconstruction efficiency at MuC3 in presence of beam-induced backgrounds, determined in thermal Higgsino events with a realistic beamspot distribution~\cite{Accettura:2023ked}. The efficiency is around 45\% for central tracks with $p_T$ between 500 MeV and 1 GeV. The null efficiency in the highest-$p_{T}$ bins is due to lack of signal events, not reconstruction effects. A conservative way to extend this map for recasting purposes would be to extrapolate the last non-null value to higher $p_T$.}
    \label{fig:trackEff}
\end{figure}

The presence of BIB affects the analysis in two ways: 1) by decreasing the tracking reconstruction efficiency $\epsilon_{\text{trk}}$ of the ST and 2) by creating random hit patterns that can be reconstructed by the tracking algorithm in our detector simulation as extra ST.
Our results indicate that for the most favourable $\theta$ region, the ST reconstruction efficiency reaches about 45\% for $p_T \geq 500$ MeV, as shown in Fig.~\ref{fig:trackEff}. The presence of BIB causes a decrease of a factor 2--4 in the ST reconstruction efficiency when compared to a background-free scenario.
Fake tracks from BIB hits mainly populate the forward regions of the tracker. 
We select tracks with at least 8 hits in the tracking detectors~\cite{Accettura:2023ked} and no more than 3 expected-but-missing hits on the fitted trajectory. STs must have a small transverse ($<3$~mm) and longitudinal ($<15$~mm) impact parameters. These requirements reduce the rate of fake tracks by more than five orders of magnitude to a negligible level. 

Additionally, BIB energy deposits in the calorimeter could be reconstructed as fake photons that can be paired with uncorrelated ST from the IP producing fake signal events. We verified using the full detector simulation described below that by appropriately raising the calorimeter cell energy thresholds the photon fake rate for photons $p_T\ge20$~GeV becomes negligible.
We expect future reconstruction techniques to achieve similar rejection power by exploiting timing information and shower shapes, and have neglected fake photons in this study.

Finally, additional sources of backgrounds could emerge from beam dynamics. For example, lepton pairs from incoherent production via two photons can be generated at the IP. The spectrum of these pairs is very soft ($p_T\le200$ MeV) so they do not contaminate the SR~\cite{schulte:2024almost}.

{\bf Analysis and Detector Simulation -} Monte Carlo samples for signal and backgrounds are simulated, targeting an effective integrated luminosity of 10~ab$^{-1}$, using \textsc{MadGraph5}\_aMC@NLO~3.5.1~\cite{Alwall:2014hca} interfaced to \textsc{Pythia}~8.307 \cite{Sjostrand:2014zea} for the parton showering and hadronisation. The response of the detector, described in~\cite{Accettura:2023ked}, is modelled using GEANT~4~\cite{GEANT4:2002zbu}. Events were reconstructed and analyzed using the \textsc{MuonColliderSoft} software within \textsc{key4hep}~\cite{Sailer:2020fah,Ganis:2021vgv}.

Tracks are reconstructed with the ACTS toolkit~\cite{Ai:2021ghi} using a combinatorial Kalman filter algorithm seeded by hit triplets. The tracking algorithm is configured with the default parameters~\cite{Accettura:2023ked}, with the exception of the minimum estimated track $p_{T}$ being lowered to 200~MeV and the width of the $\chi^2$ hit search window at each layer being increased to 30. Furthermore, in order to save computational time, no hit triplets were build outside of the barrel region of the Vertex detector, causing the sharp drop in efficiency for $\theta<0.9$ in Fig.~\ref{fig:trackEff}.

Dedicated particle gun samples including simulated BIB data from the MARS15 package~\cite{Mokhov:2017klc} for $\sqrt{s}=1.5$~TeV collisions were used to derive track reconstruction efficiency corrections that are applied to all reconstructed events when computing the final predictions. The $\sqrt{s}=1.5$~TeV BIB simulation is expected to be a pessimistic approximation based on the studies available in FLUKA~\cite{Ferrari:2005zk,Bohlen:2014buj} at $\sqrt{s}=10$~TeV.

\begin{figure}[t]
\centering
\includegraphics[width=1.02\linewidth]{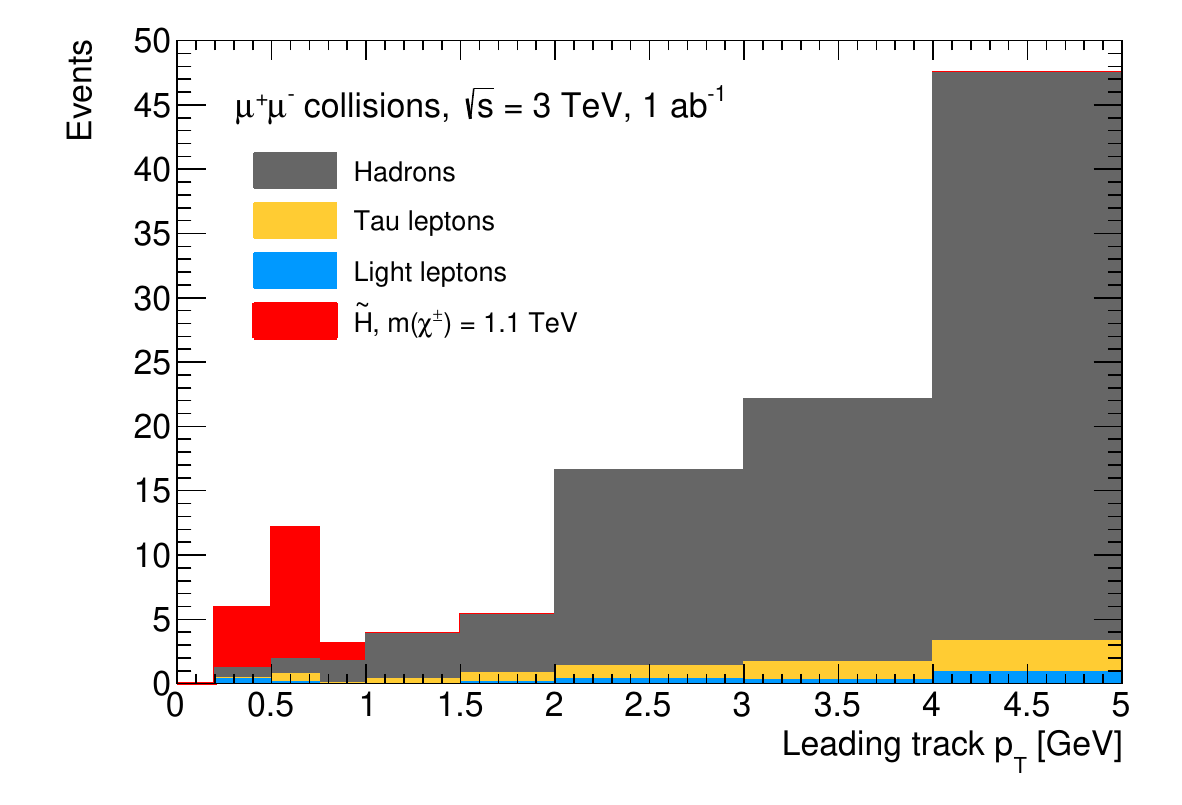}
\caption{Signal and background $p_T$ distribution of the leading ST at MuC3 for events passing all signal region selections, except that on the leading ST $p_T$. The signal, shown as a red histogram, corresponds to the thermal Higgsino-like state.}
\label{fig:leading_soft_track}
\end{figure}

{\bf Results and Discussion -} Fig.~\ref{fig:leading_soft_track} shows the $p_T$ distribution of the leading ST after our events selection at MuC3, and compare it with the main backgrounds. By selecting $p_T < 750$~MeV we can obtain a sensitivity larger than $5$ sigma for the thermal dark matter benchmark. An improved pattern recognition algorithm could substantially improve the sensitivity by increasing the ST reconstruction efficiency for a similar fake rate. This result demonstrates that in spite of the BIB, charged soft particles can be looked for with sufficient signal efficiencies in the central region of the detector. To our fortune, chargino production at the MuC is dominated by the DY process, whose angular distribution peaks at the central region, while background processes (and BIB-related tracks) peak at the forward regions of the detector. 

\begin{figure}[t]
    \centering
\includegraphics[width=1.02\linewidth]{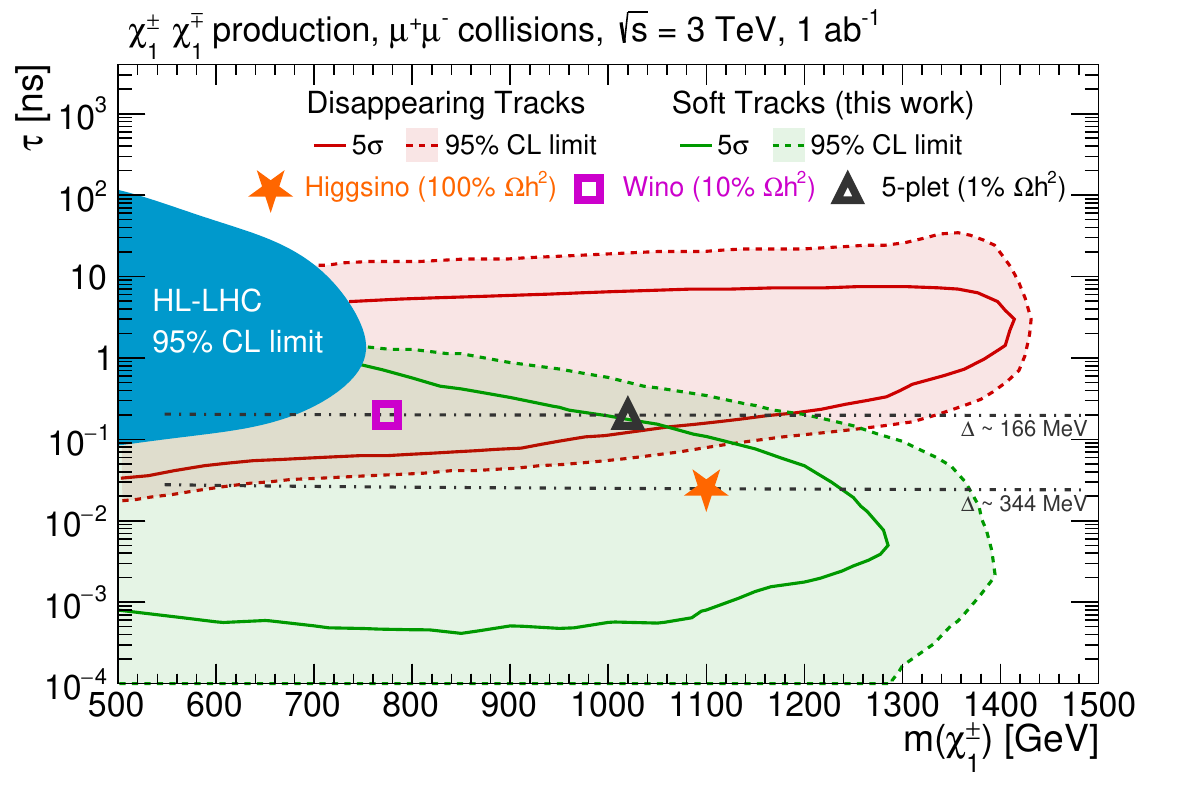}
\caption{Reach on the parameter space of MDM models. The blue region will be covered by the high luminosity stage of the LHC. Solid (dashed) red and green lines refer to 5$\sigma$~(2$\sigma$) contours from DT and ST analysis at the MuC3, respectively. The star is the Higgsino thermal target of 1.1 TeV. The square (triangle) represents a 774~GeV (1.02~TeV) Wino-like~(5-plet) state with a mass splitting of 166 MeV whose neutral component corresponds to 10\% (1\%) of the DM in the Universe.}
    \label{fig:Hino_Muc3_sensitivity}
\end{figure}

Obtaining the expected sensitivity of our ST analysis for arbitrary models is straightforward. Since the background distribution is shown in Fig.~\ref{fig:leading_soft_track}, only the signal ST $p_T$ distribution is required, which is obtainable by reweighing the $p_T$ parton level distribution with the efficiency map shown in Fig.~\ref{fig:trackEff}, and applying the selection criteria outlined in our analysis above. Following this recipe, Fig.~\ref{fig:Hino_Muc3_sensitivity} shows the expected sensitivity of a Higgsino-like state as a function of the mass and mean proper lifetime of $\chi^{\pm}$. For a given mass, increasing the lifetime means decreasing $\Delta$, which consequently modifies the chargino decay channels. We accounted for all these variations by utilizing the results in~\cite{Ibe:2023dcu} as we scanned the parameter space. Our scan includes lifetimes as low as $1.1\times10^{-4}$ ns ($\Delta\sim1.5$ GeV). Below this point, charmed-flavored decay channels start to appear and the theoretical calculation of the BR suffers from large hadronic uncertainties~\cite{Ibe:2023dcu}. Our thermal Higgsino-optimized selection criteria are applied to all masses and lifetimes, causing a loss of sensitivity for $c \tau \lesssim \sim 8\times 10^{-4}$ ns. A mass-lifetime dependent optimization, enlarging the covered parameter space, is foreseeable.

For illustration purposes, we present results from DT \cite{Capdevilla:2021fmj} and two signal benchmarks corresponding to triplet (fiveplet) MDM states with a mass splitting of 166 MeV and a mass of 774 GeV (1.02 TeV), respectively, whose neutral component account for 10\% (1\%) of the DM in the Universe \cite{Mitridate:2017izz}. This demonstrates how a combination of DT and ST analyses can reveal these MDM scenarios.

It can be confidently concluded that the soft-track analysis presented in this study is apt at successfully detecting the thermal Higgsino-like Minimal Dark Matter particle. Moreover, our technique proves instrumental in characterizing the dark sector by facilitating the measurement of mass splittings, potentially enabling the distinction of MDM from other models. It is noteworthy that our generic approach is versatile and applicable to a range of other beyond the Standard Model benchmarks, extending beyond the scope of MDM. The realization that a Muon Collider can uncover highly plausible dark matter scenarios, even in its early operational phase at an energy of 3 TeV, stands as a significant milestone for the MuC program. \\

{\bf Acknowledgments -} The authors thank Daniel Schulte and Daniele Calzolari for useful discussions, Priscilla Pani, Andrea Wulzer, Zhen Liu, Simone Pagan Griso, Donatella Lucchesi, Karol Krizka, and Karri Di Petrillo for their thorough discussion of the draft, and the International Muon Collider Collaboration (IMCC) for fostering this effort. The authors are thankful to {\it Theorist of the Month} program at DESY, which enabled the conditions for the development of this project at its early stage.
This manuscript has been authored by Fermi Research Alliance, LLC under Contract No.~DE-AC02-07CH11359 with the U.S. Department of Energy, Office of High Energy Physics. JZ is supported by the Generalitat Valenciana (Spain) through the plan GenT program (CIDEGENT/2019/068), by the Spanish Government (Agencia Estatal de Investigación) and ERDF funds from European Commission (MCIN/AEI/10.13039/501100011033, Grant No.~PID2020-114473GB-I00). This work has benefited from computing services provided by the German National Analysis Facility (NAF).

\bibliography{STatMUC}

\end{document}